\documentclass[conference]{IEEEtran}

\usepackage{subcaption}
\usepackage{hyperref}
\usepackage[usenames, dvipsnames]{color}
\usepackage{amsmath}
\usepackage[table,xcdraw]{xcolor}

\usepackage{cite}

\ifCLASSINFOpdf
   \usepackage[pdftex]{graphicx}
\else
   \usepackage[dvips]{graphicx}
\fi
\usepackage{amsmath}
\usepackage{algorithm}
\usepackage{algpseudocode}

\usepackage{array}
\usepackage{url}

\begin{document}

\title{Network Reconnaissance and Vulnerability Excavation of Secure DDS Systems}

\author{
    \IEEEauthorblockN{Ruffin White and Gianluca Caiazza and Chenxu Jiang and Xinyue Ou and Zhiyue Yang \\and Agostino Cortesi and Henrik Christensen}
        \IEEEauthorblockA{Contextual Robotics Institute\\
        UC San Diego, California, USA} 
}

\maketitle

\begin{abstract}
Data Distribution Service (DDS) is a realtime peer-to-peer protocol that serves as a scalable middleware between distributed networked systems found in many Industrial IoT domains such as automotive, medical, energy, and defense. Since the initial ratification of the standard, specifications have introduced a Security Model and Service Plugin Interface (SPI) architecture, facilitating authenticated encryption and data centric access control while preserving interoperable data exchange. However, as Secure DDS v1.1, the default plugin specifications presently exchanges digitally signed capability lists of both participants in the clear during the crypto handshake for permission attestation; thus breaching confidentiality of the context of the connection. In this work, we present an attacker model that makes use of network reconnaissance afforded by this leaked context in conjunction with formal verification and model checking to arbitrarily reason about the underlying topology and reachability of information flow, enabling targeted attacks such as selective denial of service, adversarial partitioning of the data bus, or vulnerability excavation of vendor implementations.
\end{abstract}

\begin{IEEEkeywords}
Data Distribution Service, IoT Protocol, Security, Network Reconnaissance, Formal Verification
\end{IEEEkeywords}

\section{Introduction}
\label{par:introduction}

The ubiquity of connected and autonomous devices defined as Internet of Things (IoT) and Industrial IoT (IIoT), has uncovered how the limited resources and the weak security design choices that have been made in the past represent a source of concerns in terms of safety and security in deployment. To address those problems we can distinguish between two lines of research, either studying the security and hardening solutions for the devices or focusing on the communication infrastructures. 

This work regards the latter; in particular we discuss Data Distribution Service (DDS)\cite{omg2015dds} from the Object Management Group (OMG), a widely used\footnote{\href{https://www.omgwiki.org/dds/who-is-using-dds-2}{https://omgwiki.org/dds/who-is-using-dds-2}} real-time middleware communication mechanism based on a publish-subscribe model. This standard, used in several industries including Automotive, Transportation, Healthcare, Energy systems, Aerospace, Defense, etc., permits one to build large scaled distributed networks without relying on a centralized server. However, such applications require a rigid security mechanism since any potential vulnerability can possibly lead to millions in economic losses or damages. In order to cope with requests, a design for Secure DDS Plugins \cite{omg2018ddssecure} has been developed. This enhanced version of the original DDS protocol adds authentication, access control, domain protection and cryptographic support to the standard.

\begin{figure}
    \centering
        \includegraphics[page=2,
            width=\linewidth,
            trim= 0 0 0 0,
            clip]
            {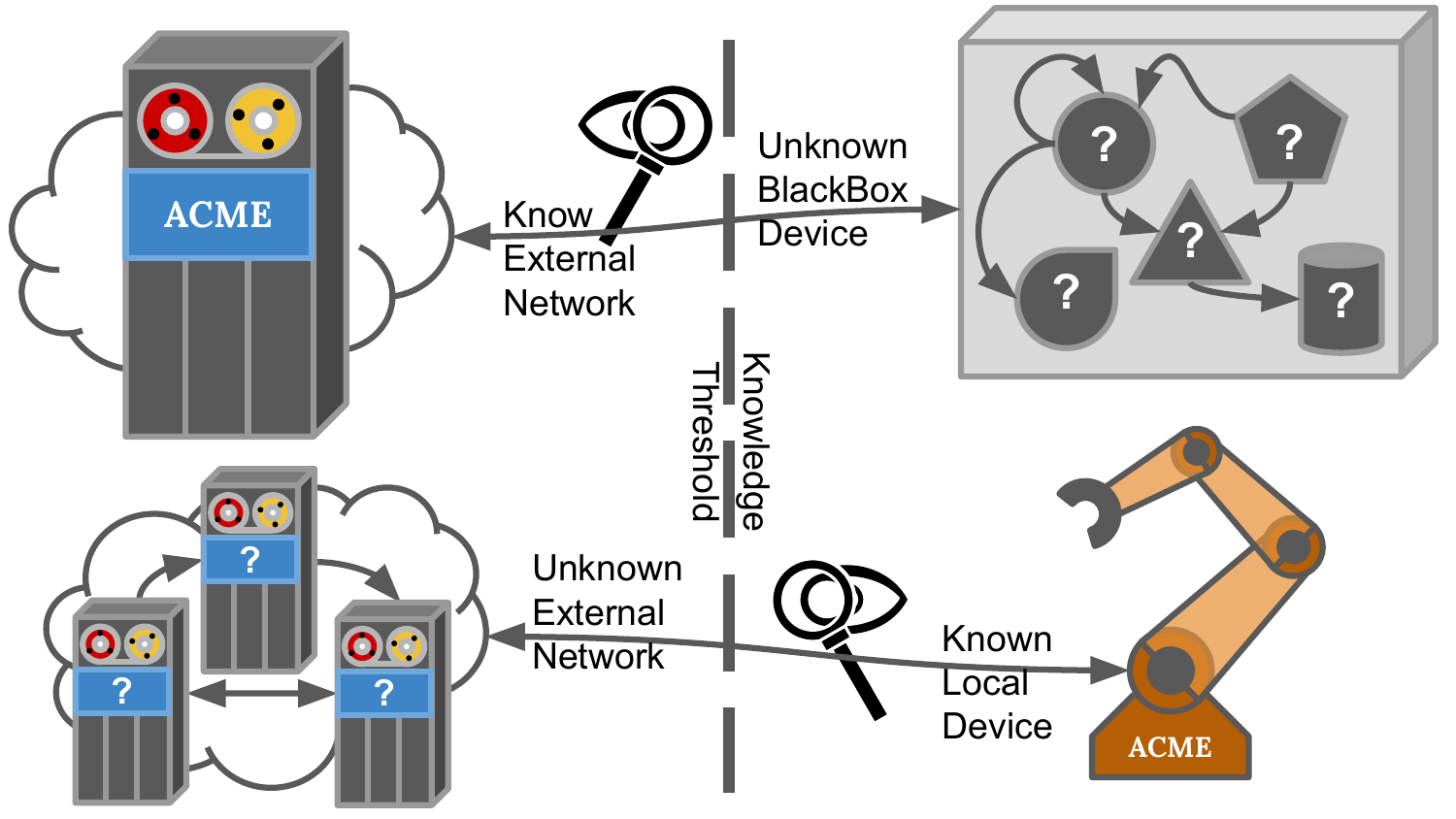}
    \caption{An example scenario where a external observer wishes to internally analyze either a closed off device or an external restricted network by monitoring traffic between observable systems. Such an adversary may be excavating for vulnerabilities, attempting to recover a system's hidden state or internal connectivity, beyond what IP packet sniffing may merely infer.}
    \label{fig:knowledge_threshold}
\end{figure}

In further detail, the security model adopted is meant to provide: confidentiality of the data samples, data and messages integrity, authentication and authorization of DDS writers and readers, message-origin and data-origin authentication, and optionally non-repudiation. By enforcing those properties, threats such as unauthorized subscriber and publisher creation, tampering and replay messages, and unauthorized access to data, are blocked.
Nevertheless, the proposed threat model doesn't cover permission confidentiality \footnote{\href{https://issues.omg.org/issues/DDSSEC12-13}{https://issues.omg.org/issues/DDSSEC12-13}}. In fact, by analyzing the plugin, we may observe how participant' handshakes are performed by exchanging a plain text permission file. Although digitally signed to preserve integrity and block an unauthorized node from accessing resources via forged permissions, its transmission plain text voids its confidentiality.
Permission files define a node's capability to read and or write data in a certain domain on the databus. By leaking such information, an attacker can infer the application layer topology by comparing the capabilities of each domain participant and deducing possible connections without having to decrypt ciphered message data. A case example, where the topic names themselves may remain sensitive, could include when a topic offers some clue as to the amount of confidential resource. If any topics are indexed sequentially, attackers can use a classical statistical theory of estimation, similar to that applied during WW2 to solve the 'German tank problem'\cite{ruggles1947empirical}; e.g. for estimating the number of surveillance sensors or alarms armed in a network or in a physical subsystem.

Additional uses for associating topics between participants is that of system identification. Higher level frameworks relying on DDS, such as the second generation Robotic Operating System (ROS2)\footnote{\href{https://index.ros.org/doc/ros2}{https://index.ros.org/doc/ros2}} often make use of standardized naming conventions, such as including the software package name or data type in the topic namespaces advertised. Fingerprinting via these clear text traits can assist in recognizing un-patched or exploitable versions of software/firmware. For example, navigation software stack or device drivers frequently include topics such as \textit{/navigation/obstacle\_range} or \textit{/bumblebee2/left/image\_raw}. Thus the significance or internal function of networked devices as with automotive ECUs or robotic sensors may remain transparent to unauthorized users, regardless of whether or not DDS Security is enabled.

Unlike traditional network reconnaissance methods like using traceroute in which an attacker needs to query the network repeatedly to obtain information about the topology\cite{mcclure2009hacking}, that may trigger alarms to network administrators, the methods we present allow an attacker to construct a richer topology of the underlying data bus merely by passively sniffing the packets inside the network.
As per the traditional case, administrators can employ techniques that obscure the network itself \cite{meier2018nethide} to impede an attacker from reconstructing the true network topology, or that trigger intrusion detection countermeasures before an actual attack is executed. In a passive attack scenario, it becomes substantially harder to identify an attacker before any malicious operation is performed. Therefore, we investigate how revealing the data flow semantics for each node and its functional role in the network renders DDS networks more vulnerable when facing malicious adversaries.

The rest of the paper is divided as follows: Section \ref{par:background} provides an overview and technical description of the secure DDS protocol and related components incorporated into our approach; Section \ref{par:threat} details our threat and attack model assumptions; Section \ref{par:approach} details our approach in partially reconstructing data bus topology and inferring reachability throughout the network at scale; Section \ref{par:implementation} documents our experimental setup and testing infrastructure; Section \ref{par:results} demonstrates how an attacker may isolate information flow from a single node by identifying critical targets or verify reachability from a selected source to target destination; Section \ref{par:related} discusses related work in relation to network reconnaissance, DDS networking and information flow control; Section \ref{par:conclusion} summarizes our main contributions and discusses potential mitigations and their caveats, as well as future work addressing remaining issues in remote access control attestation.


\section{Background}
\label{par:background}

\subsection{Data Distribution Service}
The Data Distribution Service (DDS)\cite{omg2015dds} is a standardized network middleware protocol that aims to provide reliable and scalable service based on a publish-subscribe model, i.e. a data centric model based on a conceptual \textit{Global Data Space}. The decoupled nature of publisher-subscriber compared to an ordinary request-response model renders the protocol more suitable for real-time systems and IoT applications. Applications can choose to have publishers and/or subscribers, where the data model underlying the \textit{Global Data Space}, or a DDS \textit{Domain}, is a set of data objects. A \textit{Publisher} is an object responsible for data distribution and may publish data of different data types. Similarly, a \textit{Subscriber} is an object responsible for receiving published data and making it available for the receiving application. These \textit{DomainParticipant}s can respectively write or read in a \textit{Domain}, which denotes the set of all applications that can communicate with each other. \textit{Topic} objects conceptually fit between publications and subscriptions, and uniquely identify the name, data type and corresponding Quality of Service (QoS) associated with the data on both the publisher and the subscriber sides.


\subsection{Authentication}
Each DomainParticipant must be authenticated prior to joining the DDS domain. On start, a DomainParticipant authenticates its local identity to others in the network using its own public certificate. This Identity Certificate is signed by the Identity Certificate Authority (CA)\cite{omg2018ddssecure}. Each DomainParticipant will then verify the authentication of a discovered remote peer through a mutual handshake request and reply messages. Among other tokens inside the handshake request, the Identity Certificate and the Domain Participant Permissions (detailed in next section) of a remote peer will also be included; this is precisely the information leakage we exploit in this work.

\subsection{Access Control}
In order to ensure authorization of DDS publishers and subscribers, DDS defines an Access Control Plugin. The DomainParticipant must be provisioned access to given domains, publish access to topics for data it produces, and subscribe access to topics for data it consumes. In addition, there are more configurable permissions that further segment data access, such as DDS \textit{partitions}, \textit{data tags}, and \textit{domain tags}, that are omitted from our discussion for brevity but are accounted for in our approach. Three configuration documents are associated with the Access Control Service: a Permissions CA Certificate, a Domain Governance signed by Permission CA, and a Domain Participant Permission signed by the Permission CA. The Domain Governance is a XML document specifying the protection policy inside this domain, including whether or not to enforce encryption, whether to set specific limitations on certain topics, etc. The Domain Participant Permission is a XML document containing the permissions of a DomainParticipant. Essentially, it is a set of \textit{grants} that denotes the rules to either reject or allow the DomainParticipant to write or access certain \textit{topics}, inside certain \textit{partitions} of a domain, with certain \textit{data tags} associated with the DomainParticipant. It also includes the domain the DomainParticipant allowed is to communicate in, and the time period that such permissions may be valid\cite{omg2018ddssecure}.

\subsection{Imandra}
Imandra\footnote{\href{https://www.imandra.ai}{https://www.imandra.ai}} is a formal verification tool, originally purposed for model checking financial market software and exchange protocols \cite{imandra2015create}. It is highly adaptable and performs the nonlinear arithmetic, automated induction, etc. that we need to infer the proofs or counterexamples to resolve out SAT formulation of permission intersections. Formal verification techniques can reason about a large state space without exhaustive search. Still, when surveying DDS networks at scale, solving such SAT queries would remain a bottleneck, and thus should be optimized for by reducing the number of queries required when inferring about the network. Using Imandra, we simplify the implementation of our approach by replicating the DDS Security SPI specification as functional programs in OCaml, to faithfully model default plugin logic. This also allows for generalizing our automated attack pipeline for non-default plugins; merely update the OCaml model to reflect a new SAT.

\section{threat}
\label{par:threat}

In this section, we specify both the threat and attacker models, including assumptions made when applying our approach.

\subsection{Threat Model}
In addition to information disclosure of permissions discussed, our threat model considers the following:
\begin{itemize}
    \item Network traffic may be sniffed, via live or recorded.
    \item Network topology may be originally unknown.
    \item Network semantics may be originally unknown.
    \item Network topology may be non-static.
    \item Certificate Authorities remain un-compromised.
    \item Participant issued certificates remain un-compromised.
    \item Attackers may selectively disrupt network connectivity by dropping packet traffic, route poisoning, or physically disrupting a participant device, for some non-free cost.
\end{itemize}

\subsection{Attack Model}
For the attack model, minimum requirements for execution necessitate some access to network level DDS traffic. For applications such as distributed IoT systems, a strong assumption would be that of an attacker owning all of the victim's networks simultaneously. However, neither complete nor simultaneous network access are among these minimum requirements for passive or active attacks, given that our later approach inherently reconciles with partial observability over connectivity and time. Thus multi-site measurements, such as recording IP traffic over different connections one at a time, is sufficient for reconnaissance purposes.

One may argue that access to DDS network traffic itself as a rather strong assumption for an attack scenario, given that enterprise networks often operate through VPNs. However, as DDS is a decentralized protocol supporting a range of QoS and security features, applications necessitating its adoption, often demand p2p connectivity over lossy channels that are bandwidth and energy limited. Centralized protocols dependent on reliable transport that add additional crypto overhead and deadline latency are subsequently ill-suited for these scenarios. Thus for applications using DDS, the assumption that Secure DDS traffic is observable over the physical network layer is probable, if not most likely for internal system networks, such as inside autonomous vehicles.

A representative IoT example of a highly distributed, realtime, peer-to-peer network would be the Cooperative Intelligent Transport Systems (C-ITS)\footnote{\href{https://ec.europa.eu/transport/themes/its\_en}{https://ec.europa.eu/transport/themes/its\_en}} under development of the European Commission, whose goal is to build a smart-city scaled network to exchange realtime data among vehicles and other road infrastructural facilities to optimize traffic management and take full advantage of highly automated vehicles (level 4/5)\cite{gear2017}.

We decompose the attack into two phases based on whether the attacker has the additional ability to control the network. If the attacker can only observe the network then it can perform \emph{passive} attacks; on the other hand, if it has some degree of influence on the network, \emph{active} realtime attacks become feasible.

Starting from a passive prospective, capturing permission tokens from sniffed DDS security handshake traffic enables an attacker to gradually reconstruct the underlying computation graph. In addition to mapping the structure of the computation graph to physical network topology, reconstruction of the data flow and semantic connectivity is also obtained; e.g. how devices/participants interact with each other over specific data objects.

Progressing to the active prospective, having reconstructed a rich model of system connectivity, plus some level of control over the network, an attacker is thus situated to execute far more targeted and specialized attacks. For example, if a targeted participant is directly inaccessible due to hardware protections, an active attacker may still selectively isolate it from certain data objects in the rest of the DDS domain by dropping network traffic identified as pertaining to a given topic, e.g IP port/address routes inferred from the secure handshake. In this way, an attacker may effectively remove the target participant from parts of the DDS domain, cutting the information flow from the network, or vice versa, while without overly disrupting the connectivity to the rest of the physical network, or physically compromising the host hardware.

Even if an attacker does not have the capability to control the traffic, as with secure wireless scenarios, an attacker could revert to less covert methods such as jamming the local spectrum or tampering the physical device. These methods come at significantly greater cost for the attacker, and thus the expected Return on Investment (RoI) must justify the additional risk. Again, relying on the rich model of system connectivity, an attacker may better prioritized a partial attack surface, e.g. only damaging infrastructures know to host a targeted resource or data object type for the rest of the domain.

\section{Approach}
\label{par:approach}

Under the assumptions discussed in the previous section, we know that once an attacker acquires the handshake packets it can construct the semantic network topology by interpreting the permission files. The sample snippets with Fig \ref{snippet:permissions_xml} in the appendix depict the example permission files that we will use to illustrate this process. After obtaining the network topology, we also explore how an attacker may formulate queries regarding the network's connectivity.

Example queries include: 1) Given the set of nodes (i.e. DDS participants) A and B in the cyclic graph G, what minimal set nodes in G, exclusive of A and B, would need to be disrupted to discontinue information flow from A to B; 2) Given a source A, what are the nodes that we need to offline in order to isolate all information flow from set A; 3) Given a destination set B, what set of nodes would need to be compromised to prevent B from only receiving information from the rest of G. With this information, an attacker may then selectively partition any node from the rest of the network with minimal invested effort or detectable network disturbance.

\subsection{Network Topology}
We depict the network topology as a directed graph with vertices representing nodes in the network, and edges indicating that there exists at least one topic match between the two connected vertices. The primary reason for a directed graph is that we need to distinguish publish and subscribe actions, which can naturally be described using directional edges, with edges pointing from a publisher to a subscriber. Additionally, to account for a third `relay' permission type, we decompose all relay actions to a combination of subscribe and publish capabilities on the topic. This reduction not only decreases the complexity of inferring information flow but also eases the graph visualization and introspection.

\subsection{Heuristic Graph and Lazy Evaluation}
In real world applications, a network may consist of hundreds or even thousands of nodes. Such tremendous scales inevitably make any graph construction or influencing a resource an intensive task. A naive approach to constructing a network topology requires the consideration of all permission files when computing for the potential intersect in respective permission grants. However, this is impractical given an exhaustive $O(n(n-1)/2)$ would be done using our formal verification of grant intersections; this is among the most computationally intensive steps in our attacker pipeline. Instead, our approach reduces query time latency via admissible heuristics and lazy evaluation. By first generating a heuristic graph to approximate the information flow, we substantially curtail the number of expensive inferences on grant intersections. Thus, while the initial model may exaggerate apparent connectivity, we can remain assured that resulting reachability queries via formal verification remain complete.

\begin{figure}
    \centering
        \includegraphics[page=1,
            width=\linewidth,
            trim= 0 35 0 35,
            clip]
            {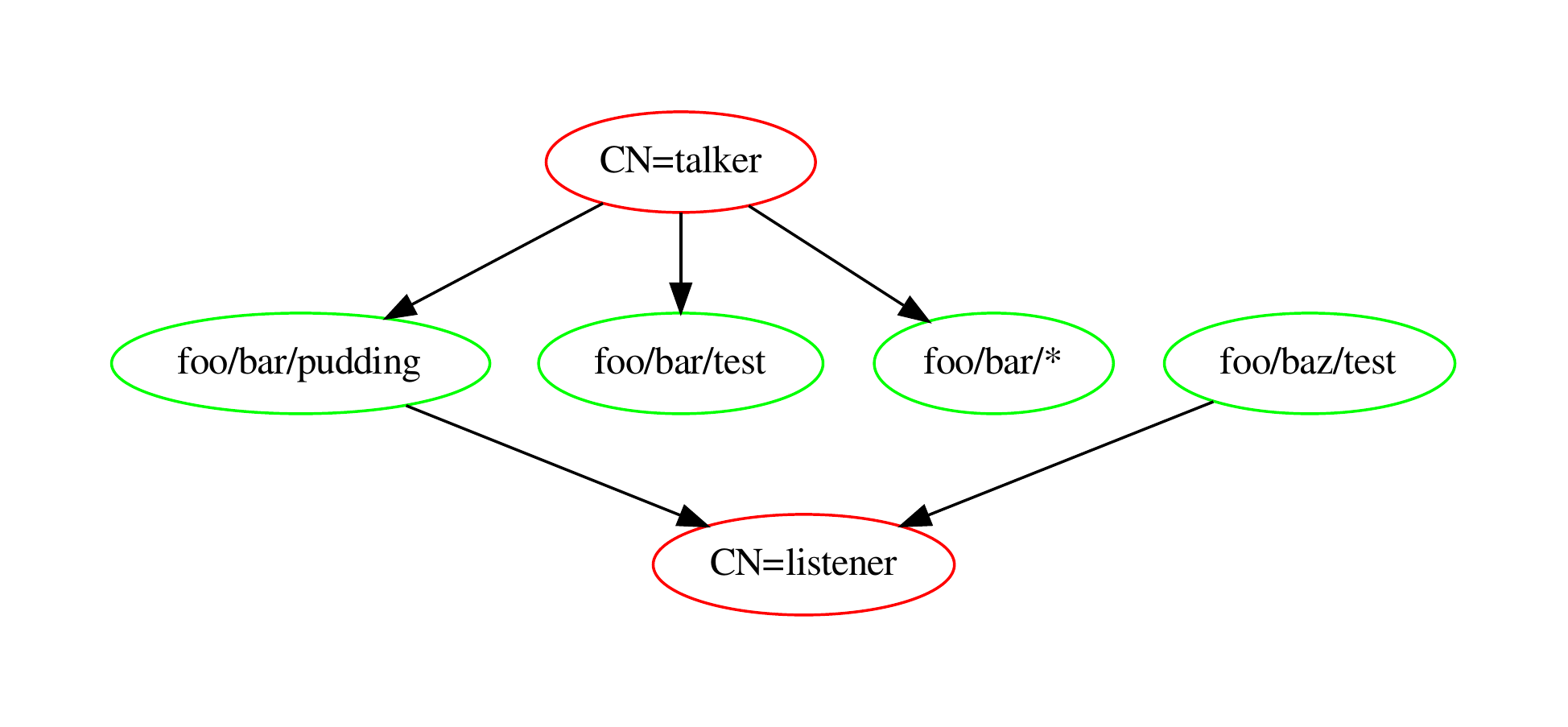}
    \caption{Raw Graph Obtained by Scanning Permission Files}
    \label{fig:raw_graph}
\end{figure}
\begin{figure}
    \centering
        \includegraphics[page=1,
            width=\linewidth,
            trim= -30 35 -30 35,
            clip]
            {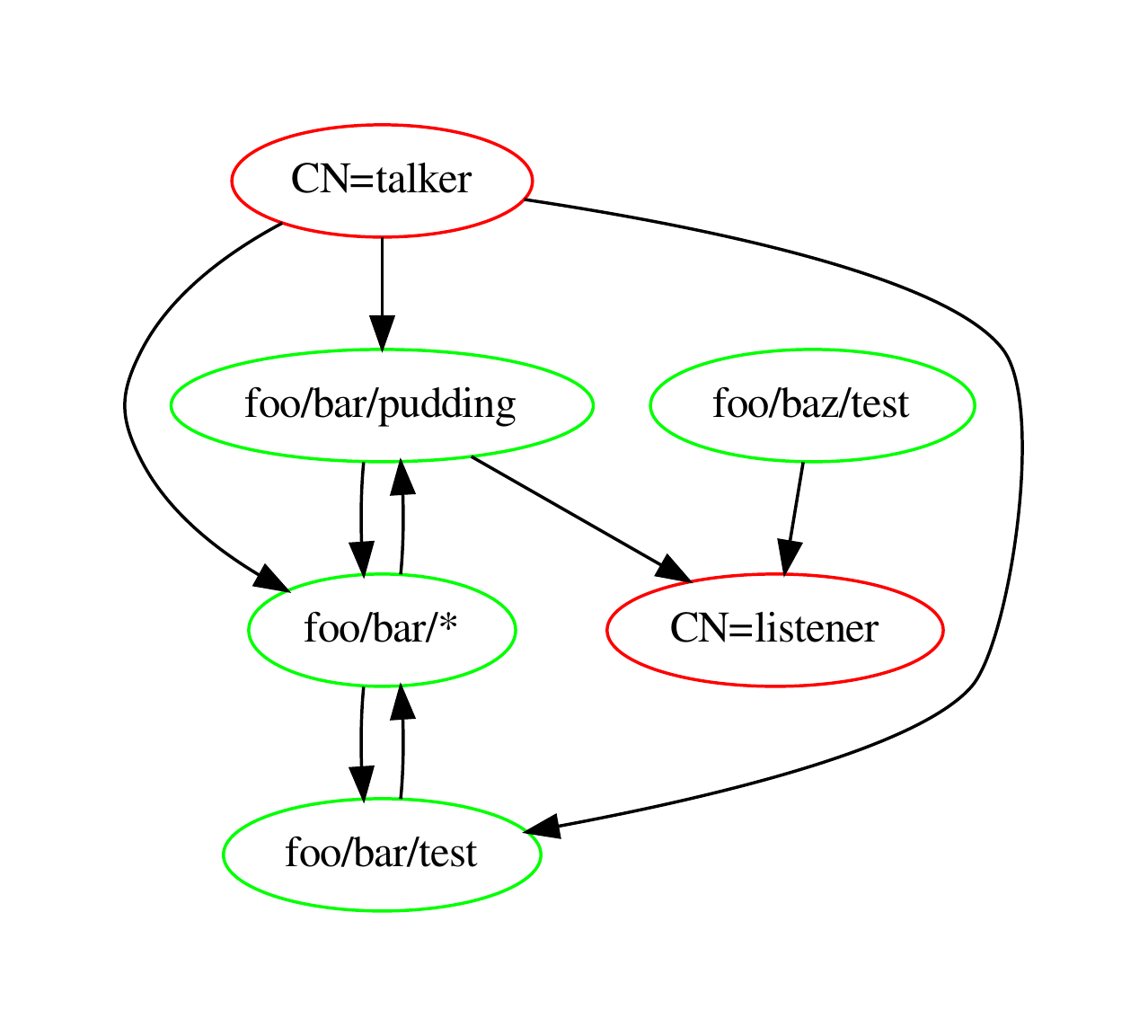}
    \caption{Connected Graph Obtained by Connecting Topics}
    \label{fig:connected_graph}
\end{figure}
\begin{figure}
    \centering
    \begin{subfigure}[b]{0.75\linewidth}
        \includegraphics[page=1,
            width=\linewidth,
            trim= 35 35 35 35,
            clip]
            {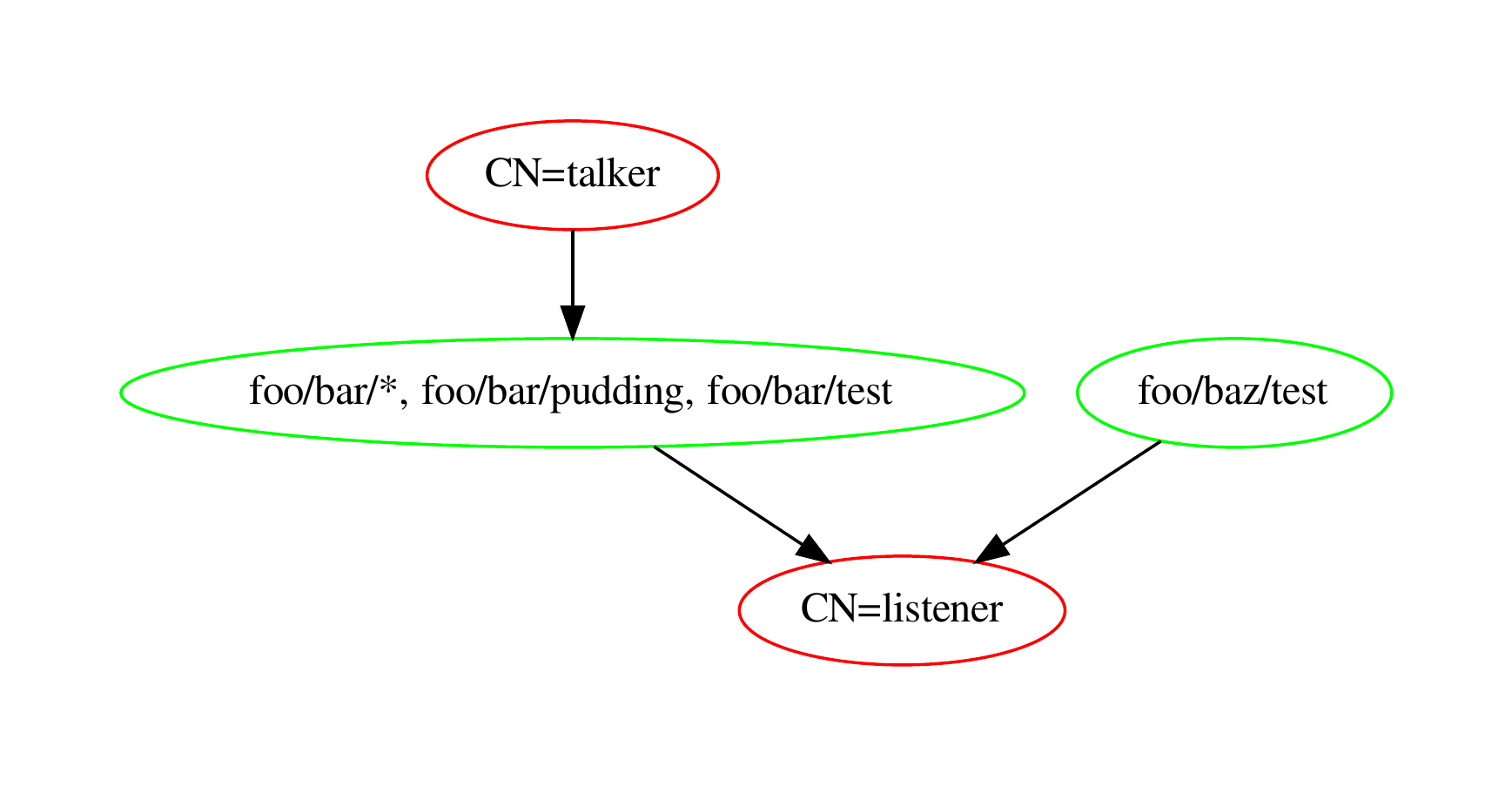}
        \caption{Contracted G}
        \label{fig:contracted_graph}
    \end{subfigure}%
    \begin{subfigure}[b]{0.25\linewidth}
        \includegraphics[page=1,
            width=\linewidth,
            trim= 35 35 35 35,
            clip]
            {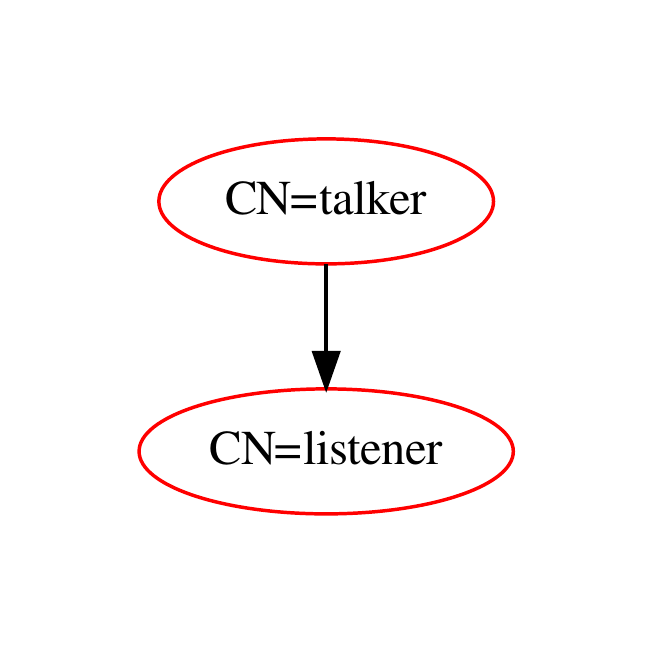}
        \caption{Heuristic G}
        \label{fig:heuristic_graph}
    \end{subfigure}
    \caption{Contracted Graph is obtained by collapsing related topics into single node, while the Heuristic Graph then is obtained by collapsing topic vertices.}
    \label{fig:reduced_graph}
\end{figure}

Generating a heuristic graph mostly relies on the fast and admissible approximation as to whether or not to connect two vertices. We decompose this approximation process into three phases: 1) First, a simple directed graph is created by indexing each grant in the permission files, adding respective vertices for both nodes as well as topics to the graph without duplicates, and then connecting nodes and topics according to the direction of information flow. This results in a directed bipartite graph such that vertex set U consists of all nodes in the network and vertex set V includes all topics involved in the network. Figure \ref{fig:raw_graph} shows a sample graph on a simple network with a talker and a listener. In this simple network, vertex set U consists of nodes talker and listener, whereas vertex set V is comprised of four topics.

This graph is quick to generate as nodes as well as topics can be iteratively appended on the fly, rather than holistically batching the entire graph all at once and performing intersection checks between any two nodes' publish and subscribe topic expressions. 

2) The second phase focuses on combining related topics to form connected components of topics and then collapses the topics into a single vertex. By combining related topics, we mean drawing bidirectional edges between any two topics that match at least once using two way `fnmatch': the POSIX string matching function chosen in the Secure DDS standard. An example of this is in Figure \ref{fig:connected_graph}, where we have one such connected component formed by three topics including \textit{foo/bar/pudding}, \textit{foo/bar/*}, and \textit{foo/bar/test}. The transition from Figure \ref{fig:connected_graph} to Figure \ref{fig:contracted_graph} illustrates the process of collapsing the connected components into a single vertex. Although this process is simple, it may potentially increase the total number of paths between different nodes in the network. The extra paths we get do not exist in the real network topology, hence a heuristic graph instead of an exact model.

3) During the last phase, we further reduce the bipartite graph to a regular network topology by eliminating topics vertex set and connect nodes that might have the capabilities to communicate on some topic. In our simple example, we get Figure \ref{fig:heuristic_graph} as a heuristic graph after completing this step, which serves as a foundation to answer the connectivity query.

Given the retrieving of a heuristic graph, naive queries on reachability using the simple edge traversal would be inaccurate; our approach resolves this via lazy evaluation. First, using the naive path computed on the heuristic graph, i.e. using Dijkstra Algorithm or A*, the resulting edge sequence or node pairs are iteratively verified for directional connectivity using a satisfaction constraint solver. We describe the reachability verification process in detail under section \ref{reachability-verification}. By pruning paths and edges sequences at query time, we avoid unnecessarily checking unfeasible flows derived from topic permission mismatches.

\subsection{Reachability Verification} \label{reachability-verification}

During the handshake phase, two DDS DomainParticipants will each verify that the other has the permission to access the resource in question. For the subject node that is advertising its access, we will abstract this into a \textit{subject} representation; containing the information about the name of the subject, the action it is requesting, the topics that it advertises to publish or subscribe, and other subject criteria regulated by access control. Algorithm \ref{fig:dds_security_plugin} in the index details how each node will validate the provided subject representation with the subject's respective permission file, and return a qualifier: ALLOW or DENY of the request.

The access control algorithm checks the grant in the permissions file that matches the supplied subject and is valid at the time it is evaluated. For this grant, it sequentially enumerates through all the rules \textit{in order}, and returns immediately if there is a match between the rule and the subject. The matching is conditioned upon many \textit{criteria}s including topics, partitions and data tags. If no rule is matched, the returned qualifier falls through to the grant's default behavior.

To check for permissive exchanges between grants and determine whether data flow between given nodes is possible, we must formally verify the intersection of the two permissions files; i.e. either assert or refute the existence of a pair of matching subjects that satisfy all pairwise constraints. More precisely, given two nodes A and B, and their corresponding permission files PermA and PermB, find two subject actions ActA and ActB such that all the following hold:

\begin{flalign}
    &Evaluate(PermA, ActA) = ALLOW \\
    &Evaluate(PermB, ActB) = ALLOW \\
    &Match(ActA, ActB)\  or \ Match(ActA, ActB)
\end{flalign}

The constraints above dictate that both subject instances must conform to the respective  
permissions, while the QoS attributes of both subjects such as topic, partition, and data tags must also correspond. The following section details the construction and consumption of such constraints.


\section{Implementation}
\label{par:implementation}

\begin{figure}
    \centering
        \includegraphics[page=4,
            width=\linewidth,
            trim= 0 165 385 0,
            clip]
        {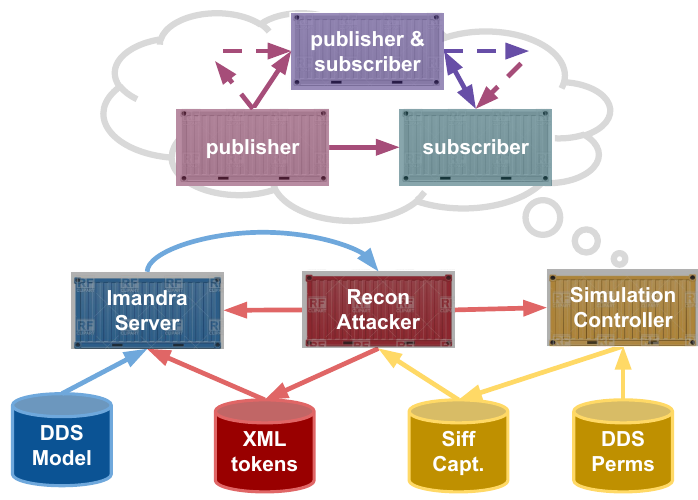}
    \caption{Visual of experimental setup and test harness. Network discovery traffic between Secure DDS participants is captured and used in concert with the SAT solver to infer application topology from intersecting permissions. The attacker then uses this feedback to precisely influence the information flow.}
    \label{fig:test_harness}
\end{figure}

To validate our approach, we construct an experimental setup with a reproducible test harness as a pipeline for the entire attacker model\footnote{\href{https://github.com/ruffsl/dds\_security\_sniffer}{https://github.com/ruffsl/dds\_security\_sniffer}}. Docker is used to containerize three main processes, as well as virtualize a targeted Secure DDS deployment, as shown in Fig \ref{fig:test_harness}.

We first programmatically synthesize a DDS application with minimal spanning permissions, valid PKI and CA trust anchors, where the digitally signed governance enforces authenticated encryption for all transport. This experimental configuration is then provided to an isolated simulation control that launches each participant in separate containers within a controlled software defined network. The first few seconds of network traffic is consecutively recorded to capture initial Real Time Publish Subscribe (RTPS) protocol discovery data, and then given to the attacker.


\begin{figure*}[ht]
    \centering
    \begin{subfigure}[b]{0.3\textwidth}
        \includegraphics[page=1,
            width=\linewidth,
            trim= 0 0 0 0,
            clip]
            {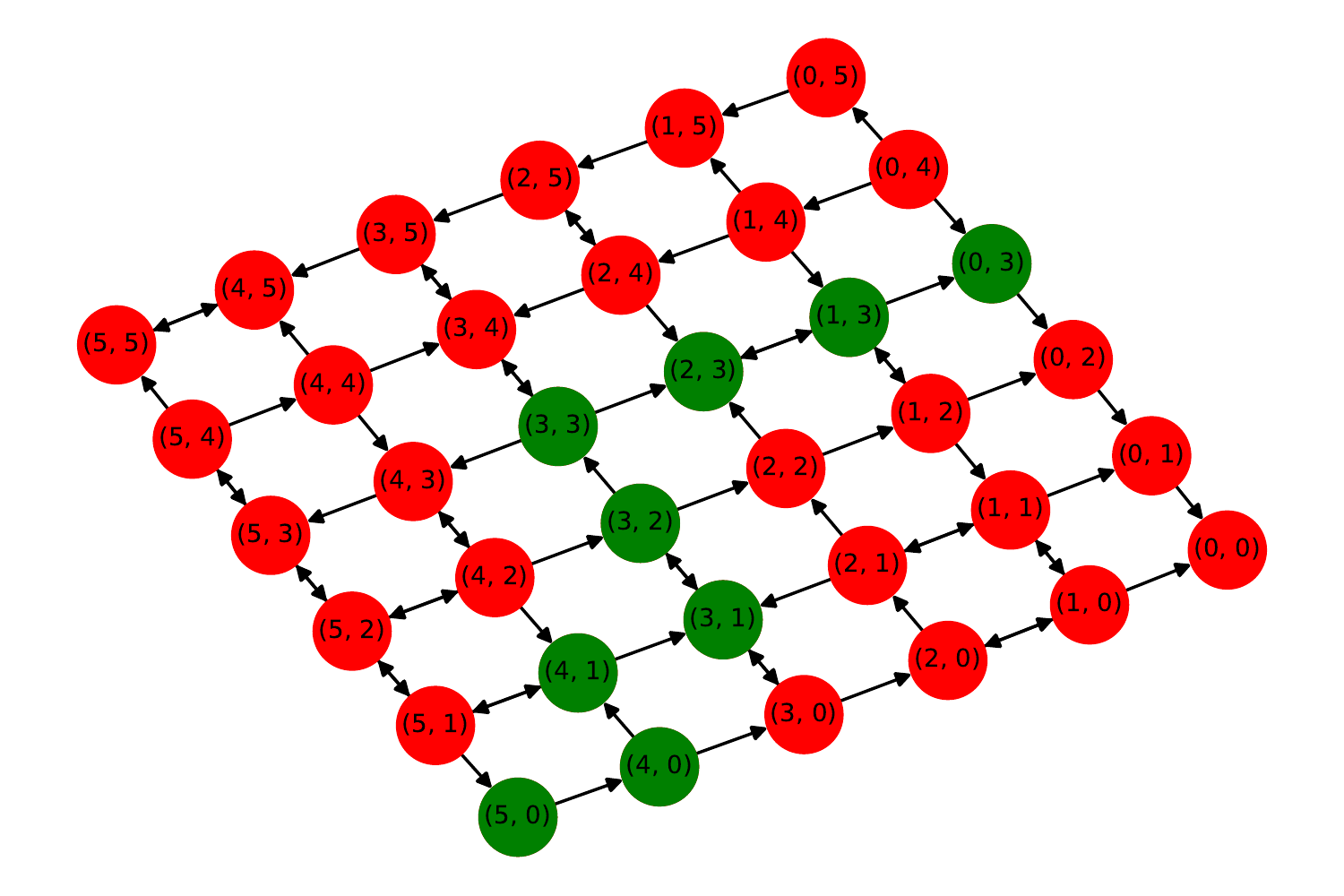}
        \caption{Prove Data Dependency}
        \label{fig:data_dependency}
    \end{subfigure}%
    \begin{subfigure}[b]{0.3\textwidth}
        \includegraphics[page=1,
            width=\linewidth,
            trim= 0 0 0 0,
            clip]
            {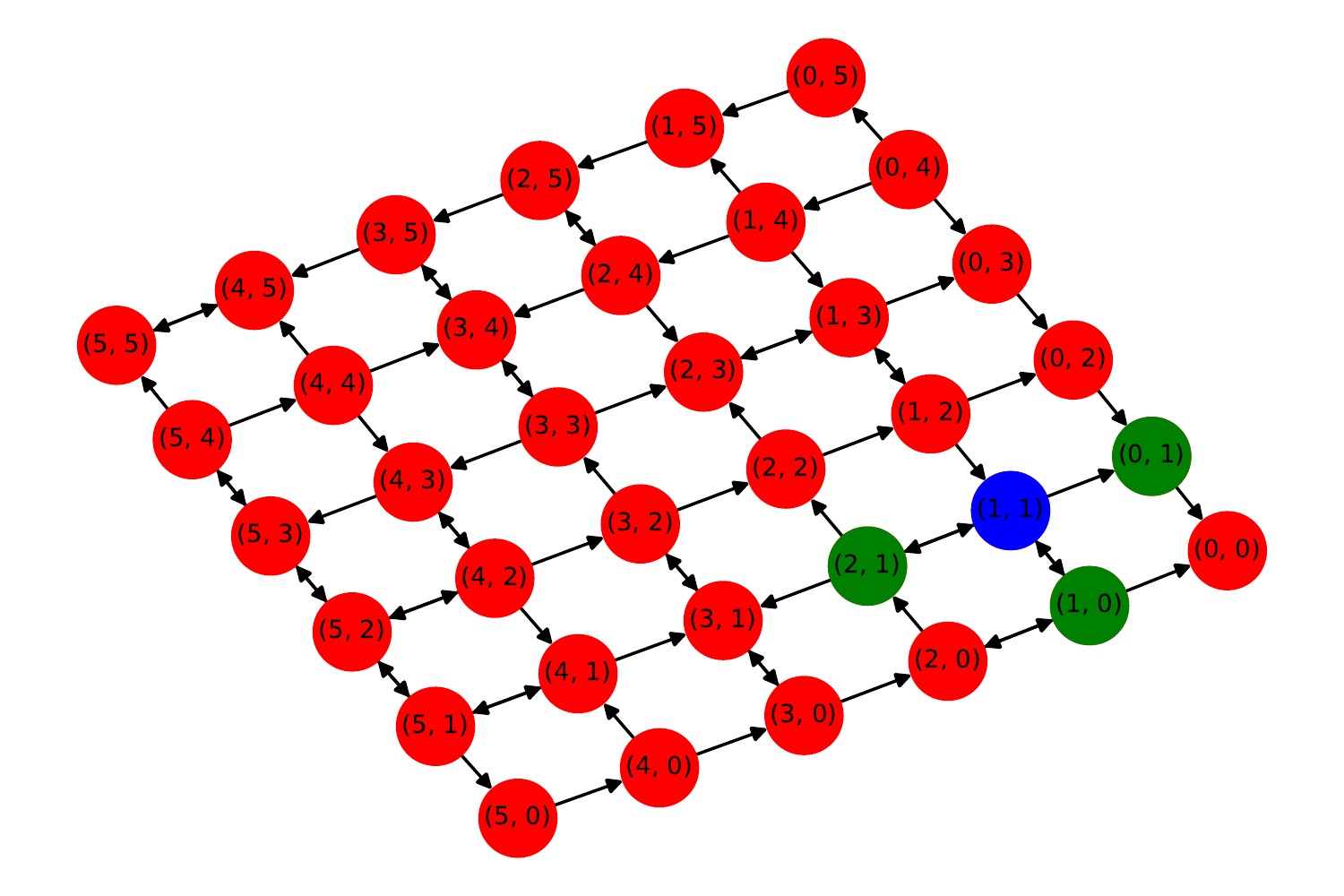}
        \caption{Isolate Publisher}
        \label{fig:isolate_publisher}
    \end{subfigure}%
    \begin{subfigure}[b]{0.3\textwidth}
        \includegraphics[page=1,
            width=\linewidth,
            trim= 0 0 0 0,
            clip]
            {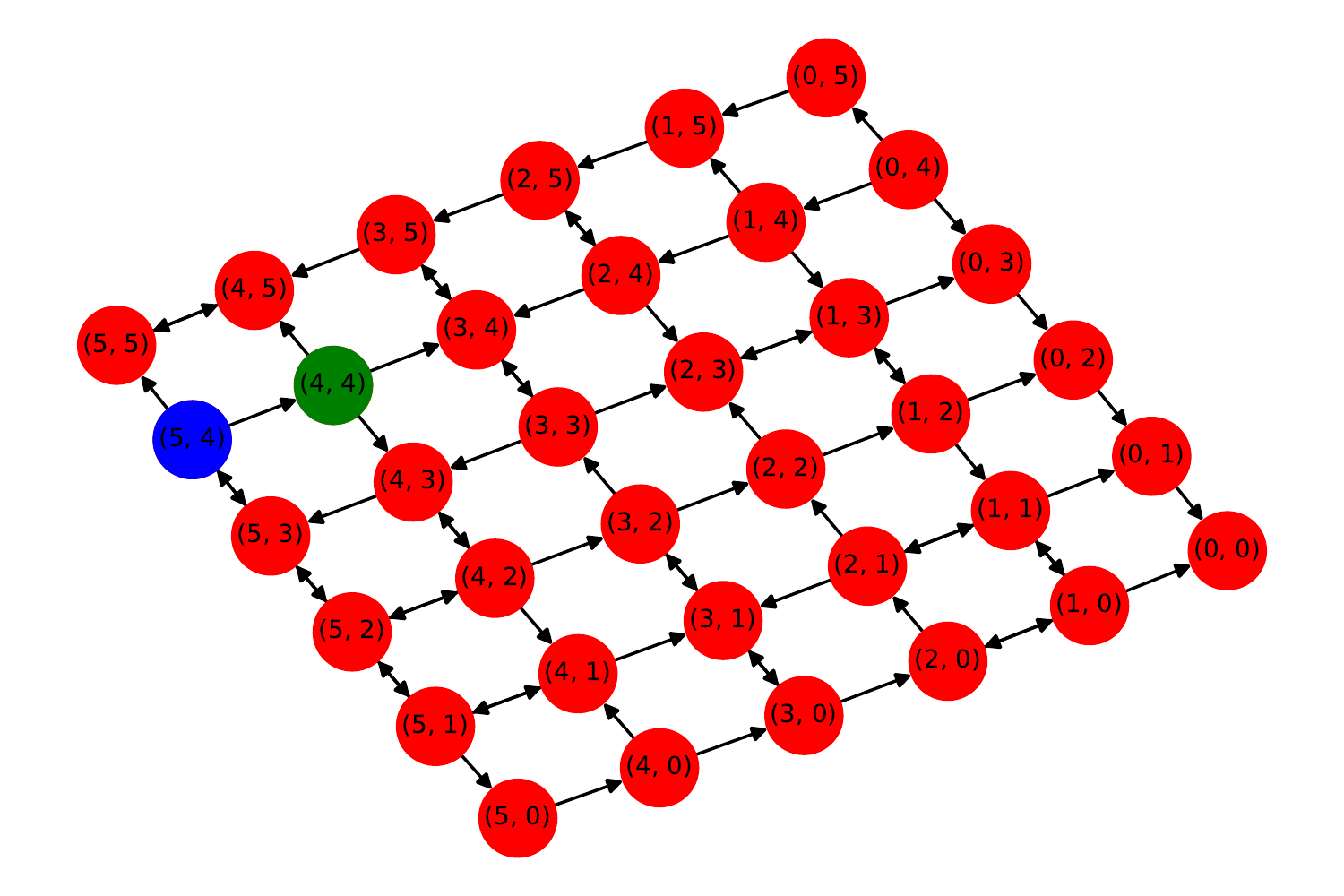}
        \caption{Isolate Subscriber}
        \label{fig:isolate_subscriber}
    \end{subfigure}
    \caption{Left to right. Query 1: given source and destination, prove data dependency. Query 2: given source, determine minimum set of nodes to isolate incoming data. Query 3: given target, determine the minimum set of nodes to cut off outgoing data.}
    \label{fig:labeled_test_graphs}
\end{figure*}

The attacker process strips all permission tokens for the raw packet capture and constructs a graph based database of permission tokens and respective origin/destination IP address. This database is then shared with the SAT solver to compute client queries.

For formal verification, we utilize Imandra as our selected SAT solver by replicating the access control evaluation logic as defined by the DDS specification in OCaml, a strongly typed functional programming language supported with Imandra. This allows us to quickly prototype and experiment with alternate security plugin designs with minimal modification.


The model of the access control logic and accompanying token database is used by the Imandra service to solve for incremental reachability inquiries from the inquisitive attacker. The attacker uses the proved or refuted subject instances as feedback to prune the heuristic graph until the overall reachability inquiry is determined. 

Armed with the associative model of DDS objects to physical network address, the attacker may finally sabotage the target application by selectively deteriorating DDS connections by commanding the simulation controller to drop specified containers from the software defined network.

To observe this disturbance, our simulated DDS network is simply composed of broadcast nodes that periodically publish a KeepAlive diagnostic message to all topics they can publish. Each node also serves as repeater, relaying any subscribed KeepAlive messages to all topics it can publish after appending its own id to avoid cyclic packets. By sampling the packet lineage from different points in the distributed application, unit tests for bisecting information flow can be verified. In the following section an illustrative example of this is presented.




\section{results}
\label{par:results}

In this section, we showcase some example scenarios to elucidate our work and demonstrate the correctness of our implementation on a more complex network; a 2D grid of consisting of 36 nodes is used to maintain readability.

\subsection{Source and Target}
If both source node and target node are specified, our model outputs a list of nodes as the path from source to target. For example, in the first subfigure of Figure \ref{fig:data_dependency}, if the source node is \textit{(5, 0)}, and the target node is \textit{(0, 3)}, then the model outputs a list of nodes containing all the green nodes.

\subsection{Source Only}
Given only a source node, our model displays a minimal set of nodes that an attacker needs to take down to prevent the source from passing data to all its subscribers. As shown in the second subfigure of Figure \ref{fig:isolate_publisher}, the source node is colored in blue, and the possible target nodes are colored in green. If the input is the blue node, then the model outputs a set including the three green nodes.

\subsection{Target Only}
Similarly, if only a target node is given, we will obtain a minimal set of nodes an attacker needs to attack to prevent the target from acquiring any new information from the network. This is illustrated in the third subfigure of Figure \ref{fig:isolate_subscriber}, where the target node is colored in green and its source node is colored in blue.

\subsection{Nonconforming Software}
Over the course of development, two notable vulnerabilities in existing DDS software where discovered while validating our default security plugin models as compared to the OMG specification verses widely used vendor implementations. Firstly, the checking of partition permissions from remote participant connections was found to been omitted from the Policy Decision Point in the access control plugin \footnote{\href{https://github.com/eProsima/Fast-RTPS/issues/443}{https://github.com/eProsima/Fast-RTPS/issues/443}}. This departure in compliance results in unintended declassification of topic data to remote participants who lack the proper authorization for participating within the same secure DDS partition.

Secondly, improper use of topic expression matching was also found in the same vendor implementation\footnote{\href{https://github.com/eProsima/Fast-RTPS/issues/441}{https://github.com/eProsima/Fast-RTPS/issues/441}}. By naively swapping arguments for the query and pattern string in the fnmatch call-sites, this allows any two participants to establish a connection using topic names with embedded expressions that match onto topic expressions lists within the permission document. This discrepancy from the specification was first observed and subsequently verified during the aforementioned experiments.

\section{Related Work}
\label{par:related}

\subsection{Network Reconnaissance} 
We have demonstrated that the permission files in clear text leak application layer topology to anyone in the same network. In fact, the encrypted packets can still leak topology information to an attacker. Other techniques are needed for us to fully defend against network reconnaissance. McClure \emph{et al.} \cite{mcclure2009hacking} presented how tools like traceroute can be used to construct the internet layer topology. An attacker may use the topology information to find the weak links in the network and DDoS attack the weakest link. To thwart reconnaissance via traceroute, Meier \emph{et al.} \cite{meier2018nethide} proposed to limit the ICMP traffic in the network or obfuscate the traceroute result. The other problem is that IoT devices usually connect wirelessly, which allows an insider attacker to eavesdrop on a large chunk of the network. If the chunk is too large, an insider attacker would be able to rebuild the application layer topology by merely examining the flow of traffic. Hakiri \emph{et al.} \cite{hakiri2015publish} proposed to connect the IoT devices with wired software defined networks using OpenFlow. Wired connections may limit the scope of possible eavesdropping and also the dynamic flexible internet layer topology nullifies the reconnaissance attempts via traceroute.

\subsection{Flow Control}
Secure DDS uses topic and partition match to enforce the flow control policy. The topic and partition expressions support fnmatch, allowing developers to build a flexible trust model. Secure DDS's label scheme is similar to the DStar labels\cite{halfond2006using} for single topic and partition but secure DDS assumes every node has the privileges to downgrade data it owns. The flow is possible as long as the subscribe set and publish set have an intersection, instead of a publish set needing to be a subset of the subscribe set. The policy opens probability not only for bad configuration but also for covert channels and allows an inside attacker to leak sensitive data. Therefore, the current flow control model works only if nodes that are granted a certificate by CA can be entirely trusted. 

\subsection{DDS} 
White \emph{et al.} \cite{White2018Procedurally} present a framework that procedurally provisions access control policies for distributed middleware. Our work extends this by adding more reachability verification on fnmatch expression to ensure that no covert channels exist in candidate policies. Khaefi \emph{et al.} \cite{Sanchez2011bloom, Khaefi2014bloom} presented how using a bloom filter in DDS node discovery phase could significantly reduce the payload of handshake traffic at the expense of a tiny chance of collision. Encoding topic discovery data into a bloom filter indeed obscures the topic expressions while providing some probabilistic integrity of the topic permissions. However, given probabilistic data structures are subject to collisions, e.g. false-positive set matches, it remains unsuitable for access control policy enforcement.

\subsection{Formal Verification for XACML}
eXtensible Access Control Markup Language (XACML) has become an attractive standard for the specification of Access Control policies given the prevalence of existing XACML tools, human and machine readable syntax, and rich set of constructs. However these same features can also make authoring XAMCL policies prone to human error. Turkmen \emph{et al.} \cite{turkmen2017formal} present a formal analysis of XACML policies by encoding them into Satisfiability Modulo Theories (SMT) formulas, facilitating formal policy analysis while relieving authors of the burden of manually proving soundness gradually. While this work remains more general in terms of access control definitions, our work additionally affords soundness checks for provisioned permissions in addition to the Policy Decision Point (PDP) logic, given that XACML is applicable to profile definitions, yet not for provisioning such profile to identities.

\section{Conclusion}
\label{par:conclusion}

In this work we introduced an approach for conducting passive network reconnaissance on systems relying upon Secure DDS, ascertaining a partial topological model of the underlying data bus, and associative mapping between data objects to network addressable participants. Using formal verification and model checking, we can then inquire about directed reachability through the distributed computation graph to efficiently perform vulnerability excavation offline without ever actively engaging with the targeted system. We then demonstrate how such acquired system models may then be used by an active attacker to prioritize targeted participants based on the data objects they represent or the connectivity they facilitate in the larger picture of the system, either by selectively isolating data flow to or from a given data producer/consumer without directly disturbing other participants. Furthermore, our methods for formal verification have been used to prove two notable vulnerabilities in existing Secure DDS vendor implementations.

Although the reconnaissance methods and vulnerability excavation tooling developed over the course of our approach may inevitably prove to be of use to malicious actors, they are also immediately beneficial for general system validation and penetration testing, as when auditing mission critical systems for flaws in access control design or implementation. For example, when certifying interface isolation between the multimedia and drive-by-wire subsystems in an autonomous automotive, manufacturers may be required to formally prove or refute the set of all satisfiable data channels between the two that would be admissible by the factory permission policy, and assure that no satisfiable channels (covert or otherwise) exist outside of the anticipated set.

The approach presented predominantly makes use of the current Secure DDS default plugin standard, thus resolving this issue would largely serve to mitigate the feasibility of the attacks demonstrated. Specifically, exchanging permission tokens in the clear during the initial crypto handshake, thus breaching confidentiality of the context of the connection is perhaps the focal issue at present. Revising the integration between the crypto and access control plugins to alternatively postponing permission token exchange through a secure channel after the crypto handshake has concluded is perhaps the most straightforward improvement. This may subsequently add another round trip delay to the overhead introduced in securing connections; however, granted the crypto handshake does not include the action request or response to begin with, it stands to reason that the permission token could be appended to the payload of the subsequent secured requests or responses.

Alternatively, one could seek to obscure the permissions embedded in the token by using an HMAC with a known key, either embedded in the token or distributing it out of band. Each topic/partition/data-tag element in the XML permission document could be replaced with say the base64 encoded digest of the expression string it replaces. Thus, upon receiving a action request from a remote participant, the local participant merely applies the same HMAC to the action and searches for the matching digests in the remote permission list. This has the benefit of obscuring permissions from those sniffing handshake network traffic while making minimal changes to existing vendor libraries. In Fast RTPS for example, the above obfuscation is implementable in less than 60 additional lines using OpenSSL. Although this works for basic string matching, support expression expansion remains an issue given the expressions in the permission list are just as obscured from the interned recipient.

However, both of these mitigations thus far, either postponing permission exchange or obfuscating the fields in the permission token have their potential drawbacks. Using HMAC is particularly vulnerable as message authentication codes only really afford integrity and not confidentiality, i.e. once an attacker knows what they are looking for, it can easily ascertain whether the permission it seeks is present in the token. Given that systems that build upon DDS, like ROS2, commonly use predictably or standardized topic names, it may be trivial to brute force obscured permissions from a limited corpus of topic names, or correlate matching digests across tokens to infer connectivity.

In postponing permission exchange, we merely delay the invocation of the Policy Decision Point, affording a secure channel to remote participants whose privileges we have not yet attested to. Only a single trusted identity need be compromised to begin scraping the permission tokens of others in the same secure distributed network. While DDS discovery information could also be decrypted with the same compromised participant identity, the permission tokens that divulge what data a participant can access versus what they currently advertise can still be advantageous to an attacker as described previously.

\section{Future Work}
\label{par:future}

A wider issue facing traditional attestation of remote privileges using digitally signed tokens is that the entire token must first be revealed in order to verify the trusted signature locally, effectively divulging all of the remote agents' capabilities, be they applicable to the current session or not. An ideal attestation method would allow a participant to prove its required privileges needed for the action at hand; nothing more, nothing less.

An alternative approach could be to fracture the token into multiple sub-tokens that are individually verifiable and only encompass a single permission. As discussed by Caiazza \emph{el al.}\cite{CCertificate}, the remote agent could then pick and choose the minimal required set of sub-tokens to be shared to gain access. This potentially adds to the complexity of the CA provisioning and expiration of permissions, as well as the coordination of exchange tokens during runtime.

This sub-token scheme would not however ultimately prevent divulging the scope of privilege for a single permission, as in the case when the permission is not just a string, but also an expression, such as a matching prefix rule for all topics starting with $/foo/*$ revealing that the remote participant also has access to $/foo/bar$.

To address this, future work could investigate the application of non-interactive zero-knowledge proofs to provide a mechanism for remote attestation of privilege in an access controlled protocol without divulging anything more than necessary. Aside from the provisioning of proving and verification key materials for PKI identities with periods of validity, particular challenges in using frameworks such as zk-SNARK \cite{sasson2013snark} (zero-knowledge succinct non-interactive argument of knowledge) with applications using DDS networks is maintaining real time performance in terms of security overhead and scalability; that is, limiting the upper bound of computation time for verification, conserving bandwidth for sending larger proofs over the wire, and limited input sizes when transforming permission sets into a boolean circuit.

\clearpage

\bibliographystyle{IEEEtran}
\bibliography{references}


\section{Appendix}
\label{par:appendix}

\begin{algorithm}
    \caption{DDS Security v1.0 Default Access Control Logic}
    \label{fig:dds_security_plugin}
    \begin{algorithmic}[1] 
        \Procedure{Evaluate}{$permissions, subject$}
            \For{$grant$ \textbf{in} $permissions$}
                \State $match \gets grant.subject\_name.match(subject)$
                \State $valid \gets grant.validity(current\_date\_time)$
                \If{$match$ \textbf{and} $valid$}
                    \State $qualifier \gets$ \Call{CheckRules}{$rules$, $subject$}
                    \If{$qualifier$ \textbf{is} $None$}
                        \State \textbf{return} $grant.default$
                    \Else
                        \State \textbf{return} $qualifier$
                    \EndIf
                \EndIf
            \EndFor
            \State \textbf{return} $ERROR$
        \EndProcedure
        \Function{CheckRules}{$rules$, $subject$}
            \For{$rule$ \textbf{in} $rules$}
                \State $domain \gets subject.domain$ \textbf{in} $rule.domainSet$
                \State $criteria \gets rule.get(subject.action.type)$
                \State \Comment{Action types: $publish, subscribe, relay$}
                \State $match \gets$ \Call{CheckCriteria}{$criteria$, $subject$}
                \If{$domain$ \textbf{and} $match$}
                    \State \textbf{return} $rules.qualifier$
                    \State \Comment{Qualifier types: $ALLOW, DENY$}
                \EndIf
            \EndFor
            \State \textbf{return} $None$
        \EndFunction
        \Function{CheckCriteria}{$criteria$, $subject$}
            \For{$criterion$, $i$ \textbf{in} $criteria.criterions$}
                \State $matches[i] \gets$ \textbf{any} $(criterion.match(subject))$
                \State \Comment{Criterion types: $topics, partitions, tags$}
            \EndFor
            \State \textbf{return} \textbf{all} $(matches)$
        \EndFunction
        \Function{Match}{$publisher$, $subscriber$}
            \State $isMatched \gets$  $publisher.action = PUBLISH$  \textbf{and} 
            \State $subscriber.action = SUBSCRIBE$  \textbf{and} 
            \State $publisher.topic = subscriber.topic$ \textbf{and} 
            \State $publisher.partition = subscriber.partition$ \textbf{and} 
            \State $publisher.data_tag = subscriber.data_tag$
            \State \textbf{return} $isMatched$
        \EndFunction
    \end{algorithmic}
\end{algorithm}

\newpage
\begin{figure*}
    \centering
    \begin{subfigure}[b]{0.5\textwidth}
        \includegraphics[page=7,
            width=\linewidth,
            trim= 0 85 455 0,
            clip]
            {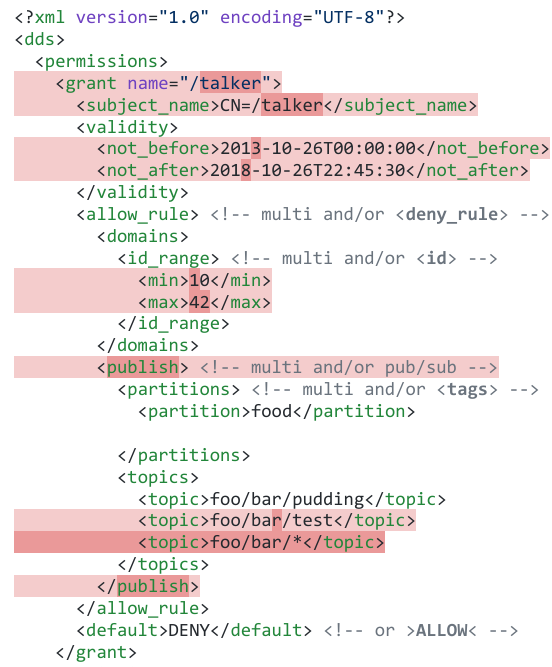}
        \caption{Talker Permissions}
    \end{subfigure}%
    \begin{subfigure}[b]{0.5\textwidth}
        \includegraphics[page=8,
            width=\linewidth,
            trim= 0 85 455 0,
            clip]
            {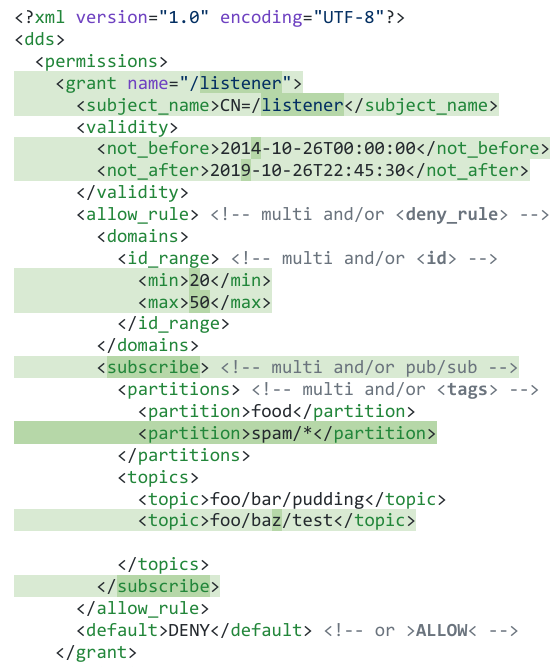}
        \caption{Listener Permissions}
    \end{subfigure}
    \caption{Highlighted diff between two Secure DDS permission.xml files depicting degrees of overlapping capabilities.}
    \label{snippet:permissions_xml}
\end{figure*}

\end{document}